\documentclass[aps,prd,showpacs,showkeys,preprintnumbers,unsortedaddress,
twocolumn
]{revtex4-1}

\usepackage{amsmath,mathrsfs,amssymb}
\usepackage{graphicx}

\begin{document}

\title{Geometrical CP violation from non-renormalisable scalar potentials}

\date{\today}

\author{Ivo de Medeiros Varzielas}
\email{ivo.de@udo.edu}
\affiliation{Fakult\"{a}t f\"{u}r Physik, Technische Universit\"{a}t
Dortmund D-44221 Dortmund, Germany}

\author{David Emmanuel-Costa}
\email{david.costa@ist.utl.pt}
\affiliation{Departamento de F\'{\i}sica and Centro de F\'{\i}sica
  Te\'orica de Part\'{\i}culas (CFTP), Instituto Superior T\'ecnico,
  Universidade T\'ecnica de Lisboa, Av. Rovisco Pais, 1049-001 Lisboa,
  Portugal}

\author{Philipp Leser}
\email{philipp.leser@tu-dortmund.de}
\affiliation{Fakult\"{a}t f\"{u}r Physik, Technische Universit\"{a}t
Dortmund D-44221 Dortmund, Germany}




\preprint{CFTP/12-005, DO-TH 12/12}

\begin{abstract}
We consider in detail the non-renormalisable scalar potential of three
Higgs doublets transforming as an irreducible triplet of $\Delta(27)$
or $\Delta(54)$. We start from a renormalisable potential that spontaneously leads to a vacuum with
CP-violating phases independent of arbitrary parameters -- geometrical
CP violation. Then we analyse to arbitrarily high order 
non-renormalisable terms that are consistent with the symmetry and we
demonstrate that inclusion of non-renormalisable terms in the potential can preserve the
geometrical CP-violating vacuum.
\end{abstract}

\maketitle


The idea that the CP symmetry is violated spontaneously
(SCPV)~\cite{Lee:1973iz,Branco:1979pv} has remarkable physical consequences. One
starts from a CP invariant Lagrangian and SCPV is achieved through meaningful
complex phases of the Higgs vacuum expectation values (VEVs) that break the
gauge symmetry group. One has further to require that no field redefinition,
compatible with the full symmetry of Lagrangian, evades all SCPV phases. SCPV
accounts for an elegant solution to the strong CP
problem~\cite{Mohapatra:1978fy, Georgi:1978xz, Barr:1979as, Nelson:1983zb,
Barr:1984qx, Kim:1986ax, Peccei:1988ci, Cheng:1987gp} and it alleviates the SUSY CP
problem~\cite{Abel:2001vy}. Also in perturbative string theory
CP asymmetry can in principle only arise spontaneously through VEVs of moduli
and matter fields~\cite{Witten:1984dg, Strominger:1985it, Dine:1992ya}. 

An interesting possibility within the framework of SCPV is when the CP phases
become calculable, so that the CP phases are independent of the Higgs potential
parameter strengths~\cite{Branco:1983tn} -- \emph{geometrical CP violation}
(GCPV). This possibility requires non-Abelian groups (for general considerations of Abelian symmetries in multi-Higgs models see e.g. \cite{Ivanov:2011ae}). GCPV was first realised by imposing the non-Abelian discrete symmetry
$\Delta(27)$~\cite{Luhn:2007uq} on the full
Lagrangian~\cite{Branco:1983tn}. GCPV was revisited 
recently~\cite{deMedeirosVarzielas:2011zw} and a new symmetry group $\Delta(54)$~\cite{Escobar:2008vc,Ishimori:2010au} leading to the same Higgs
potential was then proposed. One of major features of GCPV is the fact that the
phases of the VEVs are stable against radiative corrections
due to the presence of the non-Abelian discrete symmetry~\cite{Weinberg:1973ua,Georgi:1974au}. 

Motivated by the promising leading
order fermion mass structures presented in
Ref.~\cite{deMedeirosVarzielas:2011zw}, it turns out to be interesting to obtain viable
Yukawa structures for the lighter generations arising at the non-renormalisable
level. If one drops the requirement of renormalisability, it becomes relevant to
study whether the non-renormalisable scalar potential resulting from these
discrete groups are still compatible with GCPV. In this Letter we complete the analysis of the Higgs potential invariant under 
$\Delta(27)$ or $\Delta(54)$ that leads to GCPV by allowing higher orders scalar terms in the potential.

We use the properties of the underlying symmetry to
analyse the possible terms and classify them according to their
effect on the vacuum. 
We proceed with the analysis of both groups simultaneously. As an even number of
triplets is required to form an invariant (a consequence of their $SU(2)$
doublet nature) most of the differences between $\Delta(27)$ and $\Delta(54)$
can not manifest themselves in the scalar potential with a single triplet
representation (and its conjugate). $\Delta(54)$ has an additional generator
that swaps only two components of the triplet, and this combines any pair of
$\Delta(27)$ invariants related by that transformation into a single
$\Delta(54)$ invariant - but it will be apparent that this minor difference does
not affect our analysis of the scalar potential, as the cyclic permutation of
all three components is a generator shared by both groups.
We start by considering the renormalisable potential $V_\text{ren}$. This serves
as a brief review of the relevant results from \cite{Branco:1983tn,
deMedeirosVarzielas:2011zw} and also to establish the notation. Given the
scalars $H^i$ are $SU(2)$ doublets (the upper index denotes they transform as a
triplet of the symmetry), invariant terms are present with an equal
number of $H^i$ and $H^\dagger_i$ (the lower index denotes $H^\dagger$
transforms as the respective conjugate representation under the
symmetry).
A renormalisable potential $V_\text{ren}$ invariant under $\Delta(27)$ or $\Delta(54)$ has then the following form:
\begin{equation}
\begin{aligned}
&V_\text{ren} = H^i H_i^\dagger + (H^i H_i^\dagger) (H^j H_j^\dagger) \\
+& (H^i H_i^\dagger H^i H_i^\dagger) + c_\theta \left[ \sum_{i \neq j \neq k} (H^i)^2
H_j^\dagger H_k^\dagger + \text{h.c} \right]\,,
\end{aligned}
\end{equation}
where repeated indices denote a sum, and we have omitted the arbitrary parameters of each term except for the single phase-dependent term that is inside the square brackets.
For the analysis of phase-dependence it is convenient to parametrise the VEVs with explicit phases:
\begin{equation}
\label{eq:VEVs}
\langle H^1 \rangle=v_1\mathrm{e}^{\mathrm{i}\varphi_1}\,,\quad
\langle H^2 \rangle=v_2\mathrm{e}^{\mathrm{i}\varphi_2}\,,\quad
\langle H^3 \rangle=v_3\mathrm{e}^{\mathrm{i}\varphi_3}\,,
\end{equation}
and in particular we refer to the phase combinations displayed in $V_\text{ren}$:
\begin{equation}
\label{eq:theta}
\theta_i \equiv -2\varphi_{i} + \varphi_{j} + \varphi_{k} \,,
\end{equation}
where we have assumed $ i \neq j \neq k$.

The VEVs obtained from minimising $V_\text{ren}$ were presented in \cite{Branco:1983tn} and
confirmed in \cite{deMedeirosVarzielas:2011zw}.
Depending on the sign of $c_\theta$, we can obtain one of two classes:
\begin{subequations}
\begin{align}
\label{eq:sola}
\langle H \rangle &= \frac{v}{\sqrt{3}}\, (1, \omega, \omega^2)
\,,\\
\label{eq:solb}
\langle H \rangle &= \frac{v}{\sqrt{3}}\, (\omega^2,1,1) \,,
\end{align}
\end{subequations}
with the calculable phase $\omega \equiv \mathrm{e}^{2 \pi \mathrm{i} / 3}$. These are natural solutions for a wide range of the parameter space. Within
each class it is possible to obtain equivalent VEVs by taking cyclic permutations of the components
(e.g.~$(1,1,\omega^2)$) or by swapping the powers of $\omega$ (e.g.~$(\omega,1,1)$).


The number of terms present in the non-renormalisable potential $V$ up to a given order increases steeply with the order considered. In order to analyse the potential we rely on the fundamental properties of the symmetries and classify the large number of terms into a manageable number of categories. One important consideration is whether the equality of the magnitude of three components of eq.~(\ref{eq:sola}, \ref{eq:solb}) is perturbed by any higher order terms, i.e.~if $v_1 = v_2 = v_3$ can be maintained. In order to address this, we note that this property of the VEVs, while not guaranteed by it, is fundamentally connected to the underlying $C_3$  cyclic permutation generator contained in both symmetries considered. In this symmetry basis for the scalars, this generator forces any invariant term to be a cyclically permuting (c.p.)~combination of the 3 scalar doublets.
Starting with the phase-independent combinations, we observe that they appear only in 3 different types. The distinguishing property of these types is how many of the three components of the triplet are included in a single part of the combination. Specifically we have either $v_1^n+v_2^n+v_3^n$, $v_1^m v_2^n + v_2^m v_3^n + v_3^m v_1^n$ or $v_1^l v_2^m v_3^n + v_2^l v_3^m v_1^n +v_3^l v_1^m v_2^n$, and each of those types of combination has individual preferences for the VEVs. At renormalisable level the first two types are present: $(H^1 H_1^\dagger)^2 + (H^2 H_2^\dagger)^2 + (H^3 H_3^\dagger)^2$ and $(H^1 H_1^\dagger)(H^2 H_2^\dagger) + \text{c.p.}$. The last type first appears at order 6: $(H^1 H_1^\dagger)(H^2 H_2^\dagger)(H^3 H_3^\dagger)$. Table \ref{I_VEVs} summarises the type of VEVs that each phase-independent combination type favours, depending on the coefficient of that combination being positive or negative.

\begin{table}
\begin{ruledtabular}
\begin{tabular}{c|cc}
 & + & -  \\ \hline
$v_i^n$ & $(1,1,1)$ & $(0,0,1)$ \\ 
$v_i^m v_j^n$ & $(0,0,1)$ & $(0,1,1)$ \\
$v_1^l v_2^m v_3^n$ & $(0,0,1) / (0,1,1)$ & $(1,1,1)$ \\
\end{tabular}
\end{ruledtabular}
\caption{Types of phase-independent combinations and preferred VEVs according to
the sign of their coefficient. \label{I_VEVs}} 
\end{table}

Although a specific invariant can include more than one type of combination, the potential can be written in terms of all the allowed invariants being assigned a natural $\mathcal{O}(1)$ coefficient and the appropriate mass scale suppressions for the non-renormalisable terms. It is then always possible to rewrite it in terms of the distinct cyclic combinations, and multiplying each unique cyclic combination there is a combined coefficient that is a linear combination involving the $\mathcal{O}(1)$ coefficients of all the invariants that contain that cyclic combination and some group theoretical factors.

In order to obtain a $(0,0,1)$ or a $(1,1,1)$ VEV, ultimately the requirement turns out to be that the combined importance of terms favouring one or the other VEV is stronger. This holds even when there is a large number of terms favouring each type of VEV. At arbitrarily high orders in the scalar potential $V$, the symmetry generically predicts either a $(0,0,1)$ or $(1,1,1)$ type of VEV due to its underlying cyclic structure.
There are exceptions to this generic prediction, related with the appearance of a $(0,1,1)$ VEV or a VEV with the hybrid form $(x,y,y)$ with the ratio $x/y$ depending on the values of the combined coefficients, but we have observed that to obtain those fine-tuning of the coefficients is required.
The reason is that at each order, the $v_i^n$ type is naturally dominating (and this effect increases with the order). On the other hand, there are also more combinations of the other types, particularly the $v_1^l v_2^m v_3^n$ type which appears most frequently in invariants. Therefore in a typical situation, with similarly valued coefficients for all invariants, the sign of the combined coefficients of $v_i^n$ and $v_1^l v_2^m v_3^n$ determines the VEV, with the $v_i^m v_j^n$ terms not affecting things unless one enhances their contributions - which would be the fine-tuning we referred to previously. So to obtain either $(0,0,1)$ or $(1,1,1)$ VEVs is quite natural and there are huge regions of parameter space that lead to them.

\begin{figure}
\begin{center}
 \includegraphics[width=6 cm,keepaspectratio=true]{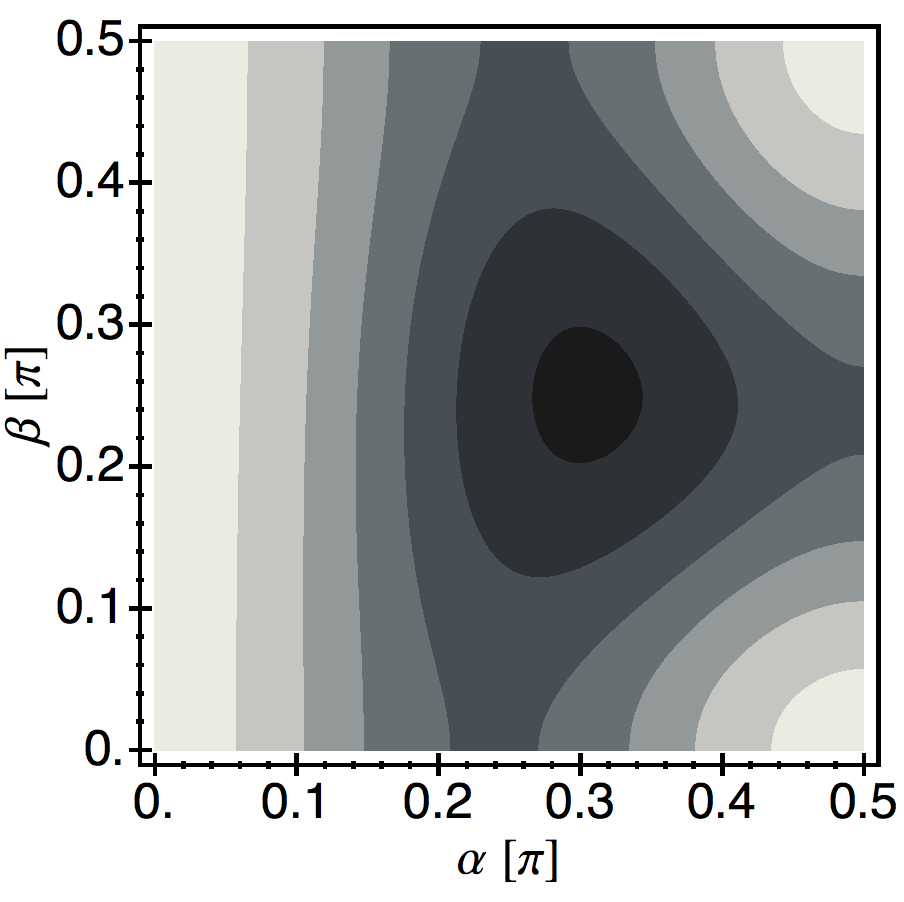}
\end{center}
\caption{\label{+-} VEV-type $(1,1,1)$ arises from cooperating terms. Note that the darker grey shades correspond to a deeper potential. The parameters $\alpha$ and $\beta$ are defined in the parametrisation given in Eq.~\eqref{eq:param}.}
\end{figure}

\begin{figure}
\begin{center}
 \includegraphics[width=6cm,keepaspectratio=true]{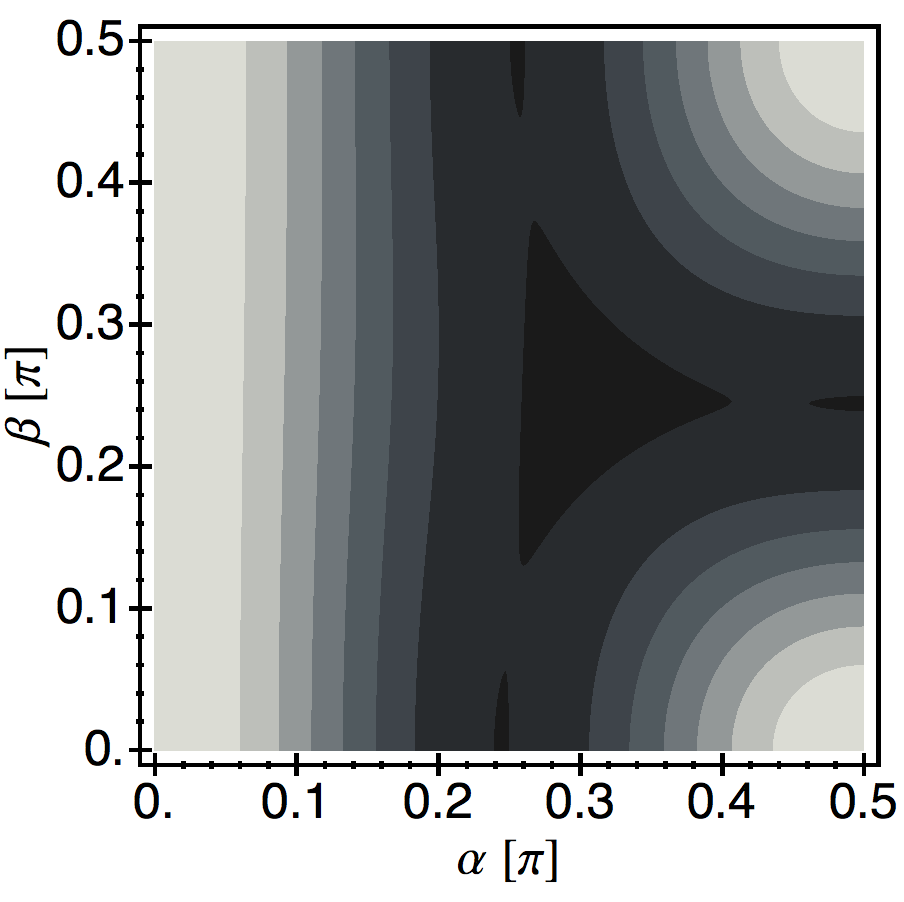}
\end{center}
\caption{ \label{++} VEV-type $(1,1,1)$ arises from dominant term. Note that the darker grey shade corresponds to a deeper potential. The parameters $\alpha$ and $\beta$ are defined in the parametrisation given in Eq.~\eqref{eq:param}.}
\end{figure}

To better illustrate this we have parametrised a VEV of constant unit magnitude, 
\begin{equation}
\label{eq:param}
\begin{aligned}
v_1&=\sin(\alpha \cdot\pi) \cos(\beta \cdot\pi)\,,\\ 
v_2&=\sin(\alpha \cdot\pi) \sin(\beta \cdot\pi)\,,\\ 
v_3 &= \cos(\alpha\cdot\pi)\,.
\end{aligned}
\end{equation}
In this parametrisation, the $(1,1,1)$ direction corresponds to $\beta = 1/4$ and $\alpha \simeq 0.30$ (strictly, $\cos (\alpha \cdot \pi) = 1/\sqrt{3}$). Due to the periodicity we focus on the region between zero and $1/2$ for $\alpha$ and $\beta$. In the case in Figure \ref{+-}, the $v_i^n$ (positive coefficient) and $v_1^l v_2^m v_3^n$ (negative coefficient) terms work together to easily produce a $(1,1,1)$ VEV. In the case in Figure \ref{++}, $v_i^n$ (positive coefficient) overpowers $v_1^l v_2^m v_3^n$ (positive coefficient) to produce a $(1,1,1)$ VEV, even though the coefficient of the $v_i^n$ is only $2/7$ of the coefficient of $v_1^l v_2^m v_3^n$. The effect of the terms $v_i^m v_j^n$ terms only becomes relevant if their coefficients are significantly enhanced. The plots shown were created for order 6, but they are representative what happens at higher orders. 
Note that in both cases reversing the signs of all the coefficients would invert the plot and would lead to the $(0,0,1)$ type of VEVs as expected.
 
We consider now the new phase dependences possible at higher orders. We once again exploit the fundamental properties of the symmetries in order to classify the large number of terms. The remaining generators shared by $\Delta(27)$ and $\Delta(54)$ are also $C_3$ factors and are fundamentally connected to the allowed phase-dependent invariants.
In \cite{deMedeirosVarzielas:2011zw} one such phase-dependence was identified: doubling the powers of the renormalisable (order 4) phase-dependent invariant produces another invariant with a distinct phase-dependence 
\begin{equation}
\sum_{i \neq j \neq k} (H^i)^4 (H_j^\dagger H_k^\dagger)^2\,.
\end{equation}
In fact this happens with any integer multiple, at a given high order new dependences $\theta^n$ are enabled
\begin{equation}
\label{eq:thetan}
\theta^n_i \equiv -2n\varphi_{i} + n\varphi_{j} + n\varphi_{k} \,,  i \neq j \neq k\,.
\end{equation}
At order 6, a distinct possibility arises:
\begin{equation}
\label{eq:eta}
\eta_i \equiv 3\varphi_{i} - 3\varphi_{j} + 0\varphi_{k} \,, i \neq j \neq k\,.
\end{equation}
It can also be generalised to integer multiples that appear at higher orders:
\begin{equation}
\label{eq:etan}
\eta^n_i \equiv 3n\varphi_{i} - 3n\varphi_{j} + 0\varphi_{k} \,,  i \neq j \neq k\,.
\end{equation}
Because of the link between the allowed phase-dependences and the generators of the groups, we can conclude that these are all the possibilities. This can be explicitly verified by computing all possible invariant products of the scalar triplet with its conjugates, and sorting out the phase-dependences. Beyond order 12 we found the number of invariants too large for this procedure to be effective, but it remains simple to verify certain properties about the $\theta^n$ combination and the $\eta^n$ combinations: they first appear through the respective powers of the lowest order terms with the $\theta$ and $\eta$ dependences, so for example $\theta^3$ and $\eta^2$ appear at order 12 respectively from $\sum_{i \neq j \neq k} (H^i)^8 (H_j^\dagger H_k^\dagger)^4$ and $\sum_{i \neq j} (H^i)^6 (H_j^\dagger)^6$.
As with the phase-independent terms discussed already, distinct invariants may include more than one type of phase-dependence, but we can rewrite the potential $V$ in terms of the unique combinations. The effective combined coefficient of each combination is a weighted sum of the $\mathcal{O}(1)$ coefficients of the invariants containing it, with group theoretical factors and the appropriate number of mass scale suppressions for the non-renormalisable invariants. As an illustration of this, in $\Delta(27)$ the product $(H \otimes H^\dagger)\otimes (H \otimes H^\dagger \otimes H^\dagger \otimes H)$ contains an invariant
$((H^1 H_3^\dagger)^3 + \text{c.p.}) + 3 ((H^1 H_3^\dagger)^2 (H^2 H_1^\dagger)+\text{c.p.}) + 3 ((H^1 H_3^\dagger)^2 (H^3 H_2^\dagger)+\text{c.p.}) + 6 H^1 H^2 H^3 H_1^\dagger H_2^\dagger H_3^\dagger$.

%

With a $(0,0,1)$ VEV the phase-dependence is lost, so from here on we consider only the $(1,1,1)$ class of VEVs.
The phase-dependent combinations also preserve the $(1,1,1)$ VEVs naturally (as a direct consequence of the non-diagonal cyclic generator). 
We can now take different combinations that share the same phase-dependence and further reduce the number of independent combined coefficients: we only need a single one for each unique phase-dependence. A demonstration of this is possible at order 6, where one can obtain the $\theta_i$ phase dependence that appears first at order 4 in two distinct ways: by combining the $\theta_1$ portion of the invariant with a matched additional $H^1 H_1^\dagger$ to obtain $[(H^1)^2 H_2^\dagger H_3^\dagger (H^1 H_1^\dagger) + \text{c.p.}] + \text{h.c.}$ or by combining the $\theta_1$ portion of the invariant with either unmatched $H^2 H_2^\dagger$ / $H^3 H_3^\dagger$, to obtain $[(H^1)^2 H_2^\dagger H_3^\dagger (H^{2,3} H_{2,3}^\dagger) + \text{c.p.}] + \text{h.c.}$.
Given a $(1,1,1)$ type of VEV, any $H^i H_i^\dagger = v^2/3$ so they all become equivalent. They are also equivalent to the already existing order 4 term with the same $\theta_i$ dependence and we can absorb their effect into a suitable redefinition of the lowest order coefficient (which is naturally dominant, given the higher order terms all have mass scale suppressions).
This procedure greatly reduces the number of relevant parameters, particularly when considering high orders where the number of invariants is huge, and allows us to treat the minimisation of the potential when a numerical approach would not be feasible.

The effect of all $\theta_i$ dependent terms is therefore already known---with a positive combined coefficient $c_\theta$ the favoured VEV is $(\omega,1,1)$, contributing $-3 c_\theta v_i^4$ to the potential, otherwise with a negative coefficient the $(1,\omega,\omega^2)$ type of VEV is favoured contributing $6 c_\theta v_i^4$ ($c_\theta <0$).

We must now consider the effect of the phase dependences that appear only at the non-renormalisable level: $\theta^n$, $\eta$ and $\eta^n$. It turns out they all preserve the existing GCPV VEVs, given suitable signs of their respective combined coefficients. Starting with $\theta^n$, we conclude for any $n$ that a positive combined coefficient $c_\theta^n$ favours the $(\omega,1,1)$ class of VEVs, contributing $- 3 c_\theta^n v_i^{4n}/M^{(4n-4)}$ to the potential. For a negative combined coefficient the $(1,\omega,\omega^2)$ class of VEVs is favoured with the potential contribution $6 c_\theta^n v_i^{4n}/M^{(4n-4)}$, where $M$ is a generic mass scale associated with the completion of the theory.
Consider next $\eta$. These terms do not distinguish the two classes of VEVs and a negative combined coefficient $c_\eta$ would preserve both classes of VEVs with a potential contribution $6 c_\eta v_i^{6}/M^{2}$. Finally, for $\eta^n$ phase-dependences the effect is the same, with negative combined coefficients $c_\eta^n$ preserving either class of VEVs with $6 c_\eta^n v_i^{6n}/M^{(6n-4)}$.


The conclusion is that it is possible to exactly preserve both the $(1,1,1)$ type of VEV together with calculable phases to an arbitrarily high order if one is willing to choose the appropriate signs of the respective combined coefficients. Note also that the $\theta^n$ or $\eta^n$ phase-dependences get a minimum of either 4 or 6 additional $v/M$ suppressions respectively.

To summarise, $\Delta(27)$ and $\Delta(54)$ are the smallest groups that lead to
geometrical complex VEVs, that violate CP symmetry spontaneously, with phases that are calculable and are stable against radiative corrections with the minimum number of three Higgs $SU(2)$ doublets.
We have investigated their non-renormalisable potentials. We described a procedure that allows to classify the possible invariants and greatly reduce the number of relevant parameters. Following this procedure we could treat the minimisation of the potential and concluded that the calculable phases can be naturally preserved to arbitrarily high order.

\section*{Acknowledgments}
The work of IdMV was supported by DFG grant PA 803/6-1 and partially through PTDC/FIS/098188/2008.
The work of DEC was partially supported by the Portuguese Funda\c{c}\~ao para a Ci\^encia e a Tecnologia (FCT) through the
projects CERN/FP/116328/2010, PTDC/FIS/098188/2008, and CFTP-FCT Unit 777
which are partially funded through POCTI (FEDER) and by Marie Curie Initial Training Network UNILHC PITN-GA-2009-237920. The work of PL was supported by the Studienstiftung des deutschen Volkes.

\bibliography{refs}

\end{document}